\documentclass[letter]{aa}
\usepackage[varg]{txfonts}
\usepackage{natbib}
\usepackage[colorlinks,linkcolor=black,citecolor=black,urlcolor=blue]{hyperref}
\usepackage{xspace}
\usepackage{graphicx}
\usepackage{afterpage}
\usepackage{multicol}
\def\caseA{ [{\rm{I}}_1\:{\rm I}_2][{\rm{BN\:x}}]}
\def\caseB{ [{ \rm{I}_1\:BN}\:][{\rm{I_2\:x}}]}
\def\caseC{ [{\rm{I_1\:x}}][{\rm{I_2\:BN}}] }
\def\outcome{[{\rm{I_1}}\:{\rm{I}_2]}{\rm{BN\:x}}}
\def\MSun{{\rm{M}}_\odot}
\def\kms{\rm{km\:s^{-1}}}
\def\bnx{\texttt{BNx-ejection}\xspace}
\def\bnxvel{\texttt{BNx-velocity}\xspace}

\begin{document}

\title{On the Formation of Runaway Stars BN and x\\ in the Orion Nebula Cluster}

\author{Juan P. Farias\inst{\ref{ch}} \and Jonathan C. Tan\inst{\ref{ch},\ref{uv}}} 

\institute{Dept. of Space, Earth \& Environment, Chalmers University of
Technology, Gothenburg, Sweden\label{ch}\mail{juan.farias@chalmers.se}
\and
Dept. of Astronomy, University of Virginia, Charlottesville, VA 22904, USA\label{uv}}

\abstract{We explore scenarios for the dynamical ejection of stars BN and x from source I
in the Kleinmann-Low nebula of the Orion Nebula Cluster (ONC), which is important for
being the closest region of massive star formation. This ejection would cause source I to
become a close binary or a merger product of two stars. We thus consider binary-binary
encounters as the mechanism to produce this event. By running a large suite of $N$-body
simulations, we find that it is nearly impossible to match the observations when using
the commonly adopted masses for the participants, especially a source I mass of
$7\:{M}_\odot$. The only way to recreate the event is if source I is more massive,
i.e., $\sim20\:{M}_\odot$. However, even in this case, the likelihood of reproducing
the observed system is low. We discuss the implications of these results for
understanding this important star-forming region. }


\maketitle

\section{Introduction}

The Kleinmann-Low (KL) Nebula is a well-studied region in the Orion Nebula Cluster (ONC),
being the closest, $\simeq$400~pc \citep{Menten2007,Kounkel2017} location where massive
stars are forming. In particular, radio source I is likely to be a massive protostar
\citep{Churchwell1987,Garay1987}.  Close to the KL Nebula is the Becklin–Neugebauer (BN)
object \citep{Becklin1967}.  BN is a young, massive
\citep[8.0-$12.6\:\MSun,$][]{Scoville1983,Rodriguez2005} star, with fast 3D motion
through the ONC of about 30 $\kms$ , i.e., it is a ``runaway'' star. The origin of this
motion has been a matter of debate.  One scenario is that BN was dynamically ejected from
the $\theta^1{\rm Ori\:C}$ system (now a binary) in the Trapezium grouping near the
center of the ONC about 4,000 years ago \citep{Tan2004}.  This hypothesis has been
supported with $N$-body simulations \citep{Chatterjee2012}, which show several current
properties of $\theta^1{\rm Ori\:C}$, including orbital binding energy and recoil proper
motion, can be understood to result from the ejection of BN.

An alternative scenario has been proposed by \cite{Bally2005} and \cite{Rodriguez2005}
who suggested that dynamical interaction of BN, source I and perhaps an additional
member, originally proposed to be radio source n, could have resulted in the high proper
motions of BN and radio source I that are approximately in opposite directions. Details
of this third member are crucial for this scenario since momentum conservation using BN
and source I alone results in a mass for source I of $\sim20\:\MSun$ in contrast to the
$7\:\MSun$ estimations from gas motions near the source
\citep{Matthews2010,Hirota2014,Plambeck2016}.

Recent observations using multi-epoch high resolution near-IR images from the
\emph{Hubble Space Telescope (HST)} \citep{Luhman2017} have shown high proper motion of
another star, source x, that strongly indicate that it was the third member of the
multiple system (see Figure \ref{fig:positions}). Given source x's mass ($\sim3\:\MSun$)
and proper motion, now the mass estimation for source I via momentum conservation and
from circumstellar disk gas dynamics are in better agreement at $\sim7\:\MSun$. It has
been also argued that if source I was a loose binary that merged during the interaction,
e.g., 6 and 1$\:\MSun$ stars, the released potential energy would be more than enough to
explain the kinetic energy of the system.

However, there are some aspects of this scenario that appear questionable. In particular,
it involves the most massive star, BN, being ejected as a single star from a binary of
two much lower mass stars, i.e., with total mass of $\sim7\:\MSun$. Thus in this paper we
carry out numerical experiments to explore this scenario. We focus on the case where a
binary source I (with components $\rm{I_1}$ and $\rm{I_2}$) interacted with another
binary composed of BN and source x in a bound system that resulted in the dynamical
ejection of source x and BN.

We present a set of $\sim10^7$ pure $N$-body scattering simulations
focused on the possible binary-binary interaction event that formed
the observed system. We first test the scenario presented by
\cite{Luhman2017} and then modify some of the parameters, especially
source I's mass, to test the sensitivity of the results.  We describe
our methods and initial conditions in \S\ref{sec:setup}, present our
results in comparison to the observed system in \S\ref{sec:results1}
and discuss our findings and draw conclusions in
\S\ref{sec:discusion}.

\section{Methods}\label{sec:setup}

We explore the scenario in which the ejection of BN and source x was
caused by a dynamical decay of a multiple system that included source
I, which was two stars in the past that may have merged as a result of
the dynamical interaction. {Ignoring situations with pre-existing
  triples and higher-order multiples that require large numbers of
  parameters for their description}, three possible cases can be
considered as initial conditions for this event involving four
members. Case 1: a binary-binary interaction; Case 2: a binary system
perturbed by two single stars; and Case 3: all stars were single
stars.  For simplicity we explore only the first case, which is
arguably the most probable since it involves close interaction of only
two initially independent systems. {Cases 2 and 3 involve the
  coordinated close encounter of 3 and 4 systems, respectively, which
  makes them intrinsically less likely. Thus, we do not consider Cases
  2 and 3 further in this letter.}

\begin{figure}
        \includegraphics[width=\columnwidth]{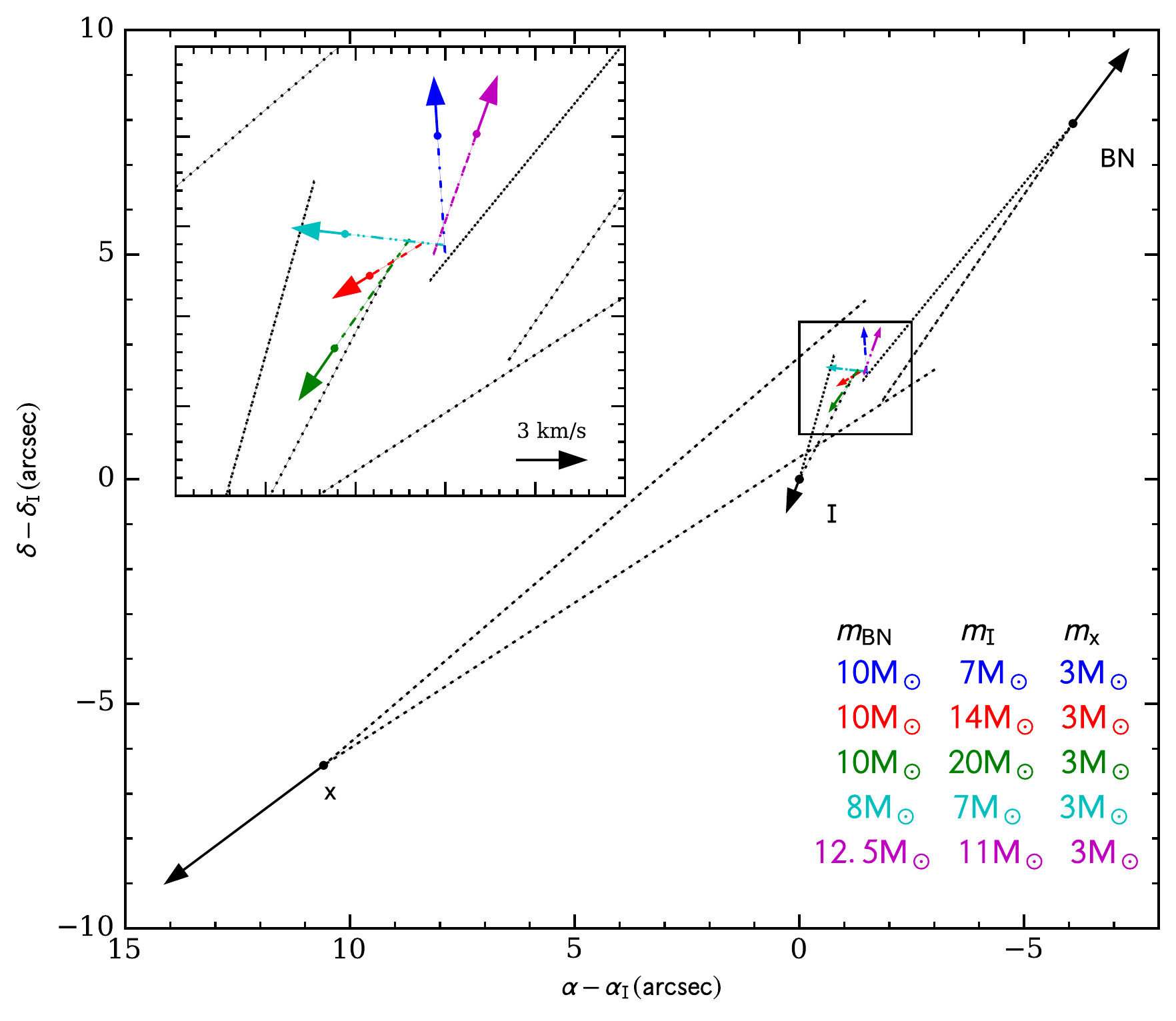} \caption{
Overview of the ejection scenario, also showing the center of mass motion given different
combinations of masses explored in this work.  Filled circles show actual positions of
the stars and of the center of mass, dashed lines track the positions 540 years ago ($1\sigma$ error cones shown for BN, I and x) and
solid lines with arrows show 100 years into the future {based on current proper motions in the rest frame of Orion \citep[see][]{Luhman2017}. Different
colors show various mass combinations of the stars that we have explored.} Positions are relative to source I
\protect\citep[$\alpha(J2000)=-05^h35^m14\rlap.^s516$ and
$\delta(J2000)=-05^\circ22'30\rlap.{''}59$,][]{Rodriguez2017}.}\label{fig:positions}
\end{figure}

Within Case 1 there are three possible initial combinations that we
label as A: $\caseA$; B: $\caseB$ and C: $\caseC$, {where [a b]
  indicates a binary pairing of stars a and b}. We look for
interactions that result in the outcome $\outcome$, i.e., with the
ejection of BN and x leaving the binary $[{\rm{I_1\:I_2}}]$ with or
without a merger, that we will refer to as \bnx. From this subset of
cases we identify those in which the velocities of the individual
stars are within 2-$\sigma$ of the observed values reported by
\cite{Luhman2017} and \cite{Rodriguez2017} as \bnxvel.

{For our fiducial case,} we adopt the same masses discussed by \cite{Luhman2017}, i.e.,
$m_{\rm{x}}=3\:\MSun$, $m_{\rm{BN}}=10\:\MSun$ and
$m_{\rm{I}}=7\:\MSun$ \citep{Matthews2010,Hirota2014,Plambeck2016}.
Assuming that source I was two stars, binary or not, we assume a mass
ratio $q=0.166$ for its members (i.e., $m_{\rm{I_1}}=6\:\MSun$ and
$m_{\rm{I_2}}=1\:\MSun$), but we have also tested a range of other
values finding no major change in the results due to this choice.  The
radius of the individual stars are taken from stellar models developed
by \cite{Hurley2000}.  We assume I$_1$ and I$_2$ are protostars 
{or pre-main-equence stars} and thus increase their radius by a
factor $\eta\ge1$ to account for the more extended radii that a
protostar should have relative to a main sequence star of the same
mass, adopting $\eta=2$ as a simple, fiducial choice.  We also test
the sensitivity of our results to this factor {finding no major
  difference on the results except when this becomes $\geq 3$ at which
  point the energy of ejections falls considerably.}

Given the previous assumptions, there are then several combinations of parameters that
set the initial conditions of each experiment.  Our standard procedure is to choose them
randomly from expected distributions, summarized as follows: (1) The semimajor axis $a$
of each binary is taken from a uniform, random distribution in logarithmic space in the
range $a=0.1-6300\:$AU. (2) The eccentricity of each binary is chosen using two extreme
distributions: A) Using only circular orbits, i.e., $e_{\rm{i}}=0$, which might be
expected if binaries formed via circumstellar disks; B) A thermal distribution
\citep{Heggie2003}, i.e., $dF_{\rm{b}}/de=2e$, which is the extreme scenario in which
binary systems have had enough time to thermalize via stellar encounters. (3) The
direction of the angular momentum vector of each binary is chosen randomly, as is (4) the
initial orbital phase of the binaries. The above parameters define the internal
properties of each binary. 

Next come the parameters that define the interaction itself. We setup the experiments in
order to only have initially bound systems, i.e., if both binaries were single stars they
would remain bound after the interaction. Therefore, (5) the relative velocity at
infinity $v_{\rm{i}}$ is drawn from a Maxwell-Boltzmann velocity distribution with
$\sigma=3\:\kms$ truncated at the critical velocity \begin{eqnarray}
        v_{c}&=&\sqrt{\frac{G}{\mu}\left(\frac{m_{11}m_{12}}{a_1}+\frac{m_{21}m_{22}}{a_2}\right)},
\end{eqnarray} where $G$ is the gravitational constant, $m_1=m_{11}+m_{12}$ and
$m_2=m_{21}+m_{22}$ are the masses of each binary, summing their respective components,
and $\mu=(m_1+m_2)/(m_1m_2)$ is the reduced mass of the system
\citep[see][]{Gualandris2004}. Thus $v_c$ is the velocity below which the total energy of
the system in the four body center of mass is negative, and therefore the ejected stars
are the result of dynamical interaction and not of the initial conditions. Also, full
ionization is not possible, i.e., there will be always a binary (or merged stars) left
behind.

Next is: (6) the impact parameter, $b$, which is drawn randomly in discrete bins of radii
$b_{i}=2^{i/2}b_0$ following the method of \cite{McMillan1996} to calculate cross
sections of the relevant interactions.  We choose $b_0=100\:{\rm{AU}}$ and increase $i$
until no relevant outcomes are encountered. Then, the contribution of the events in each
bin $i$ to the final cross section of this event $\Sigma_{\rm{X}}$ is
$\pi(b_{i}^2-b_{i-1})N_{{\rm{X}},i}/N_i$ with $N_{{\rm{X}},i}$ and $N_i$ being the number
of events X and the number of trials respectively, both inside the $i$-th bin.  The
contribution of bin $i$ to the squared uncertainty in the calculation,
$(\delta\Sigma_{\rm{X}})^2$, is $[\pi(b_{i}^2-b_{i-1})/N_i]^2N_{{\rm{X}},i}$
\citep{McMillan1996}.  For the first bin we have chosen $N_{i=1}=500\,000$.

Simulations are performed using the \texttt{Fewbody} software \citep{Fregeau2004}, an
accurate Runge-Kutta integrator which conserves energy and angular momentum to the order
of $10^{-8}$. It also uses the ``sticky star'' approximation for collisions with no mass
loss and an expansion factor of the merger product of $f_{\rm{exp}}=2$.

The above method is repeated for different combinations of the member masses. All these
combinations with their respective total momentum vectors are shown in Figure
\ref{fig:positions}.

\section{Results} \label{sec:results1}

\begin{center}
\begin{table*}
        \centering
        \caption{Interaction cross sections for the different mass combinations}
        \begin{tabular}{ccccccccc}\hline\hline
        &  &  &  &   \multicolumn{2}{c}{\bnx} &\multicolumn{2}{c}{\bnxvel} &  \\
        Case &
        Eccentricity &
        $ m_{\rm I}\:[\MSun] $ & 
        $ m_{\rm BN}\:[\MSun] $ &
        $ \Sigma\:[\times 10^6{\rm AU^2}] $ &
         BR $[\times10^{-3}]$ &
        $ \Sigma\:[{\rm AU^2}] $ &
         BR $[\times10^{-6}$] &
        $ N_{\rm sims} [\times10^6] $\\ \hline\\[-1.5ex]

A & circular & 7  & 10    & $2.82  \pm 0.02$  & 35.0& $ 0.2 \pm 0.2 $ &0.392   & 5.10 \\
A & thermal  & 7  & 10    & $1.84  \pm 0.01$  & 29.4& $   2 \pm 2   $ &1.17    & 5.11 \\ 
A & circular & 14 & 10    & $6.65  \pm 0.04$  & 39.4& $  16 \pm 8   $ &6.85    & 5.11 \\
A & thermal  & 14 & 10    & $4.55  \pm 0.03$  & 32.9& $  11 \pm 4   $ &7.88    & 5.71 \\
A & circular & 20 & 10    & $9.95  \pm 0.07$  & 38.3& $  49 \pm 9   $ &31.8    & 6.09 \\
A & thermal  & 20 & 10    & $6.72  \pm 0.04$  & 38.4& $  44 \pm 10  $ &25.1    & 5.71 \\
A & circular & 7  & 8     & $3.29  \pm 0.02$  & 37.5& $ 0.5 \pm 0.3 $ &0.780   & 5.13 \\
A & thermal  & 7  & 8     & $2.01  \pm 0.02$  & 35.2& $ 1.0 \pm 0.9 $ &1.04    & 4.81 \\
A & circular & 7  & 12.5  & $2.37  \pm 0.02$  & 28.1& $<0.06        $ &$<0.12 $& 8.50 \\
A & thermal  & 11 & 12.5  & $1.50  \pm 0.01$  & 25.5& $ 0.8 \pm 0.8 $ &0.195   & 5.12 \\ \hline \\[-1.5ex]
B & circular & 7  & 10    & $0.018 \pm 0.002$ &0.240& $0.06 \pm 0.06$ & 0.125  &  8.0 \\
B & thermal  & 7  & 10    & $0.008 \pm 0.001$ &0.208& $<0.06        $ &$<0.13 $&  7.5 \\
B & circular & 14 & 10    & $0.21  \pm 0.01 $ &4.53 & $ 1.5 \pm 0.4 $ & 2.58   &  5.4 \\
B & thermal  & 14 & 10    & $0.101 \pm 0.004$ &3.87 & $ 2.1 \pm 0.6 $ & 3.12   & 4.81 \\
B & circular & 20 & 10    & $0.53  \pm 0.01 $ &12.7 & $  25 \pm 14  $ & 13.7   & 5.41 \\
B & thermal  & 20 & 10    & $0.293 \pm 0.008$ &10.4 & $  15 \pm 2   $ & 16.8   & 5.12 \\
B & circular & 7  & 8     & $0.023 \pm 0.001$ &0.458& $0.06 \pm 0.06$ & 0.133  & 7.51 \\
B & thermal  & 7  & 8     & $0.011 \pm 0.001$ &0.347& $ 0.2 \pm 0.2 $ & 0.444  & 4.5  \\
B & circular & 11 & 12.5  & $0.018 \pm 0.003$ &0.124& $<0.06        $ &$<0.20$ & 5.12 \\
B & thermal  & 11 & 12.5  & $0.0069\pm0.0008$ &0.111& $<0.06        $ &$<0.13$ & 7.40 \\ \hline \\[-1.5ex]
C & circular & 7  &  10   & $0.017 \pm 0.001$ &0.490& $ 0.1 \pm 0.1 $ &0.125   & 8.01 \\
C & thermal  & 7  &  10   & $0.011 \pm 0.001$ &0.273& $ 0.06\pm 0.06$ &0.125   & 8.01 \\
C & circular & 14 &  10   & $0.045 \pm 0.002$ &1.35 & $ 3   \pm 2   $ &0.823   & 8.51 \\
C & thermal  & 14 &  10   & $0.043 \pm 0.004$ &1.34 & $ 4   \pm 1   $ &3.13    & 5.12 \\
C & circular & 20 &  10   & $0.083 \pm 0.004$ &2.15 & $ 51  \pm 18  $ &8.21    & 5.12 \\
C & thermal  & 20 &  10   & $0.081 \pm 0.005$ &2.66 & $ 36  \pm 16  $ &8.67    & 5.42 \\
C & circular & 7  &  8    & $0.020 \pm 0.001$ &0.847& $ 0.1 \pm 0.1 $ &0.133   & 7.51 \\
C & thermal  & 7  &  8    & $0.012 \pm 0.001$ &0.494& $ 0.13\pm 0.09$ &0.266   & 7.51 \\
C & circular & 11 & 12.5  & $0.011 \pm 0.001$ &0.281& $ 0.1 \pm 0.1 $ &0.222   & 4.51 \\
C & thermal  & 11 & 12.5  & $0.011 \pm 0.002$ &0.141& $<0.06        $ &$<0.20 $& 5.11 \\ \hline\hline
\end{tabular}
\label{tab:results}
\end{table*}
\end{center}

\begin{figure*}
        \centering
        \includegraphics[width=\textwidth]{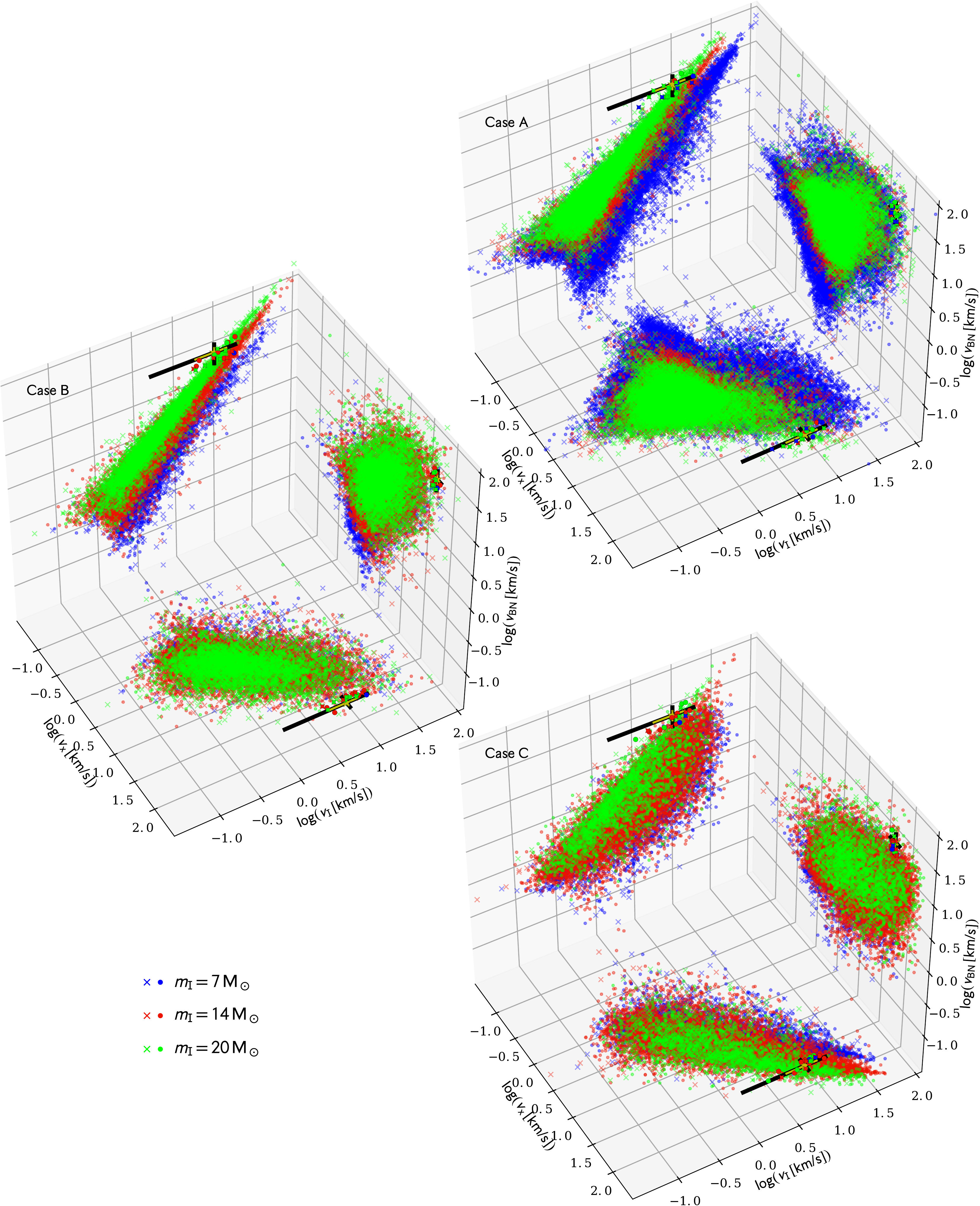}\\
        \caption{
Simulation results compared with observed velocities. Case A, Case B
and Case C panels shows the resulting velocities that match the \bnx
event in the three velocity planes for sources I, x and BN for each
respective initial combination varying only the mass of source I.
Crosses and filled circles represent the adopted eccentricity
distribution with thermal and circular eccentricities, respectively.
Yellow star and errorbars shows the observed values with its standard
error.  Highlighted symbols represent \bnxvel matches. Black errorbars
shows the range on which \bnxvel was searched, i.e., $2\sigma$
errors.}
        \label{fig:scatter}
\end{figure*}
Table \ref{tab:results} summarises the resulting interaction cross
sections {and branching ratios (BR), i.e. the number of cases over the total number of simulations for each configuration,} for the \bnx and \bnxvel cases. The interaction cross
sections of the \bnx case are considerably larger for Case A, with
$\Sigma_{\bnx}=(2.8$ and 1.8) $\times10^6{\rm{AU}}^2$ for the fiducial {(i.e., $m_{\rm BN}=10\:M_\odot, m_{\rm I}=7\:M_\odot, m_{\rm x}=3\:M_\odot$)}
circular and thermal cases, respectively.  Such large cross sections,
together with branching ratios of several percent of the \bnx event in
the fiducial case that we show in Figure \ref{fig:br} {and discuss
  further in Appendix \ref{sec:br}}, implies that the ejection of the
massive BN object and source x from the system is quite possible.

However, when considering the velocities of the ejected stars in the
\bnxvel case, the cross sections drop to $\sim1\:{\rm{AU}}^2$ in all
fiducial cases.  We have checked that this result is independent of
the assumed mass ratio {of the source I components}. Also, if
$\eta$ is greater, then the chance of obtaining the observed
velocities becomes even smaller, due to the energy constraints that
the radii of the stars imply. We have found that to match the observed
velocities the parameter $\eta$ must be no larger than 3, i.e., for a
given mass, the protostars should not have radii larger than 3 times
the radii of a main sequence star of the same mass.

The situation becomes more favorable only if the mass of source I is
$>7\:\MSun$. In the best case we explored, with
$m_{\rm{I}}=20\:\MSun$ cross sections increase by a factor between
$\sim20$ to 600 in the different initial configurations.  However, the
cross sectional areas are still small, $\sim$10 to 50 AU$^2$, which
means that these events are quite rare {when compared to the whole
  ensemble of outcomes.}  In Appendix \ref{sec:br} we present and discuss
the branching ratios of all the possible outcomes of our experiments
sorted from the most to least probable.

\afterpage{
\begin{figure}
        \centering
        \includegraphics[width=\columnwidth]{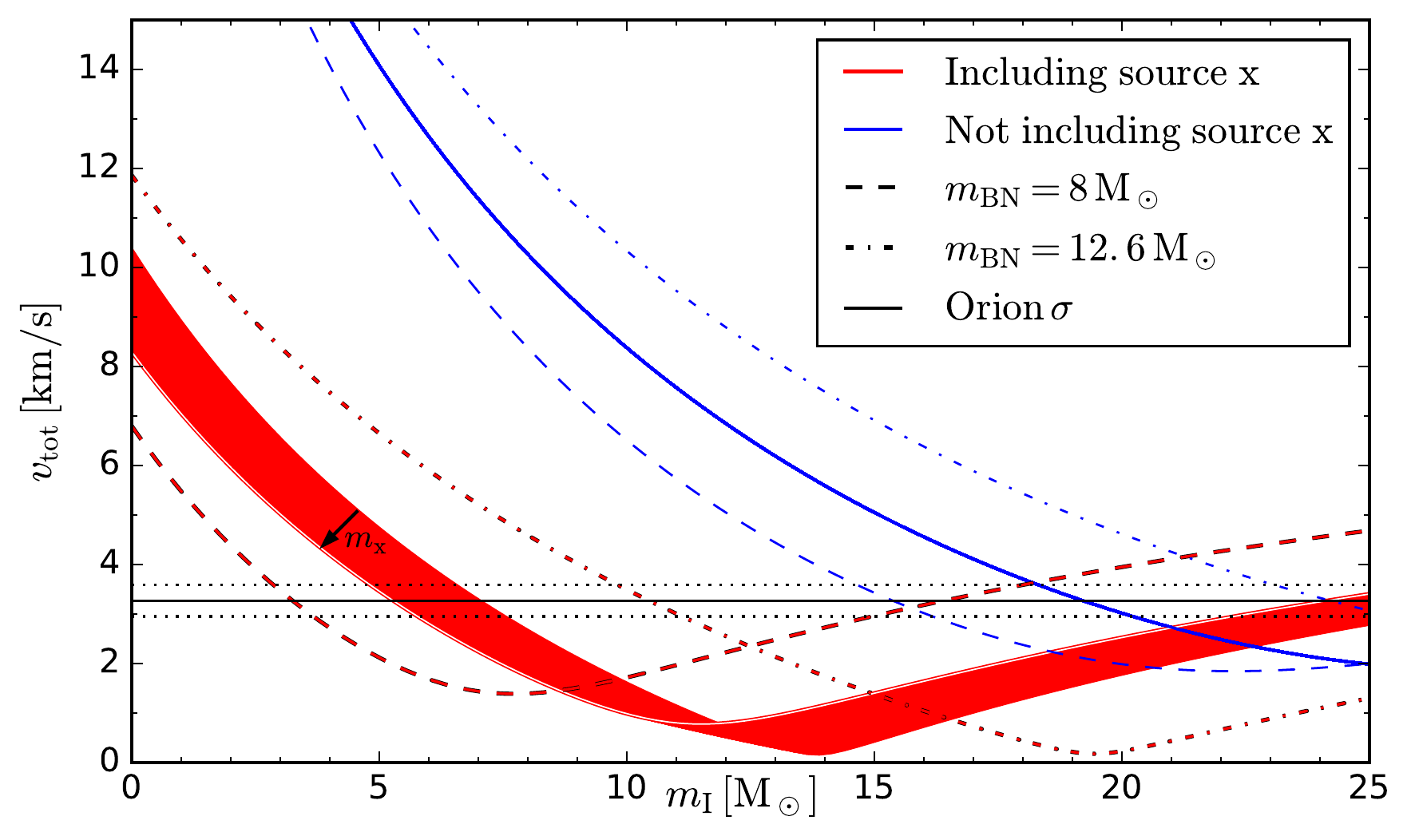}
        \label{fig:momentum}
        \caption{
The system (center of mass) velocity as a function of the mass of
source I (see also Figure \ref{fig:positions}). Red lines show the
scenario where source x is part of the ejection event as suggested by
\cite{Luhman2017} \protect\footnotemark.  The red shaded area shows
the result of varying the mass of source x between 2.5-3.0$\:\MSun$,
with the black arrows showing the direction of the increment.
Blue lines show the scenario if source x was not part of the event. In both
red and blue cases, the dashed and dot-dashed lines show results of
adopting the lower and upper limits of BN's mass, respectively
\citep{Rodriguez2005}. }
\end{figure}
\footnotetext{
Note that there is an error in the units for the velocity associated with the specific momentum that they mention: it should be $1.4\:\rm{mas\:yr^{-1}}$ instead of
$\kms$.}}

\section{Discussion and Conclusions}\label{sec:discusion}

Our simulations indicate that a more massive source I has a better
chance to produce the observed system. Figure \ref{fig:scatter} shows
a scatter plot of the velocities obtained in each setup when varying
source I's mass. The yellow star and errorbars show the observed
system. There is a very specific trend that appears in the
$v_{\rm{BN}}-v_{\rm{I}}$ panels in Figure \ref{fig:scatter} showing
that the observed system appears to be one order of magnitude above
the trend for source I that has a mass of $7\:\MSun$. Increasing
source I's mass naturally places simulations in agreement with
observations. Even though we have shown this type of event is quite rare {in terms of branching ratios with respect to all possible outcomes}, if
by chance it happens with the correct released energy, the observed
velocities would be only achieved if the mass of source I is not so
small. This is also supported by the other velocity panels in all the
Cases A, B and C.

{Some caveats associated with our analysis should be
  mentioned. For example, we have ignored the dynamical effects of gas
  expulsion. The modeling of \cite{Chernoff1982} implies about
  $4\:M_\odot$ of gas has been ejected from the central region over a
  period of 1200 years requiring an energy of
  $\gtrsim5\times10^{47}\:$erg. If a large fraction of this gas was
  ejected impulsively at the time of dynamical interaction, then this
  could alter some of the specific results of our analysis: e.g., an even greater energy needs to be liberated in the dynamical interaction. However,
  we note that the gas has been ejected quasi-isotropically from BN-KL
  \citep[see, e.g.,][]{Allen1993,Bally2017}, so that the effects on the
  plane-of-sky momentum analysis are not expected to be so
  large. Still, future modeling could allow for, e.g., a variable mass
  of stars during the interaction, i.e., sudden mass loss occuring as
  part of any merger event. Other caveats include that we have ignored
  the dynamical effects of any other masses, including other
  surrounding stars and gas components. Still, these are expected to
  be relatively minor since the velocities of the stars are relatively
  large compared to the velocity dispersion of the ambient material in
  the region.}

The recent evidence that source x was involved in the ejection of
the BN Object appeared to reconcile the incongruence between the mass
of source I estimated by momentum analysis and estimations via
rotation of its putative circumstellar disk. However, if we re-do
momentum analysis with the new data we can see that source x's
participation is not enough support for the low mass estimations of
source I. Figure \ref{fig:momentum} shows the momentum analysis for
the old and new scenarios as a function of source I mass. Blue lines
shows the scenario where only source I and BN participated in the
ejection with the upper and lower limits of BN's mass as dashed
lines. The only way the system center of mass can be moving within the
velocity dispersion of the ONC (black solid line) is if source I's
mass is at least $20\:\MSun$. Now, it has been argued that the
inclusion of source x would remove this constraint. The former is true
as we can see in the red lines on Figure \ref{fig:momentum}, however,
for the system to be moving within the velocity dispersion of the ONC,
source I's mass could actually be in a very wide range of masses, from
5 to $25\:\MSun$ with a minimum near $14\:\MSun$.

Figure \ref{fig:positions} shows the present center of mass with
different mass combinations used in this work. If the individual
masses are those adopted by \cite{Luhman2017} (blue arrow) the system
is moving mostly outwards, away from the center of the ONC.  However,
if source I is more massive (red and green arrows) the system velocity
points towards the center of the cluster, i.e., towards the
Trapezium. We consider that a scenario involving infall of a dense
molecular gas core from which the protostars are forming is more
likely than one involving motion of the core out from the cluster
center. For example, passage near the strong ionizing radiation from
$\theta^1$C is expected to have had potentially very disruptive
effects on the core if it had previously been located near the
Trapezium.

In conclusion, we have shown that the ejection of BN and x from source
I (as a binary or merged binary) as presented by \cite{Luhman2017},
{i.e., with a relatively low mass for source I of $\sim7\:M_\odot$
  that is less than BN's mass}, is in general a very unlikely
event. In particular, with the given masses and observed velocities it
is nearly impossible to reproduce the observations {with the
  binary-binary interactions we have considered}. If the interaction
occurred as we have explored, then it is more probable that source I
is much more massive than the preferred value of $7\:\MSun$ presented
by \cite{Luhman2017} and others. Thus future measurements of the mass of
source I are needed to better constrain this proposed ejection scenario.

However, other possible initial combinations remain to be explored,
e.g., a binary perturbed by two stars (i.e., effectively a 3-body
initial interaction), or all initial single stars (i.e., a 4-body
initial interaction), but these are expected to be inherently rarer
and there is no reason to think these combinations could increase the
chances significantly since the binary-binary interaction is the most
likely to release the necessary energy to produce the ejection. {A
  single star interacting with a pre-existing triple remains a
  possibility that needs to be considered, especially if the dynamical
  mass of source I does turn out to be at the low end of the range
  modeled, i.e., $\sim7\:M_\odot$. Such interactions could also
  include unbound fly-bys. This would be the only way to reconcile the
  scenario of BN's ejection from $\theta^1C$ \citep{Tan2004,Chatterjee2012} with ejection
  of source x, although plane-of-sky momentum conservation would appear to place
  challenging constraints on such a model if BN suffered only minor accelerations and
  course deflections in such a fly-by}.


\onecolumn
\begin{appendix}

\section{Branching ratios for all the possible outcomes}\label{sec:br}

\begin{figure*}
        \centering
        \includegraphics[width=0.90\textwidth]{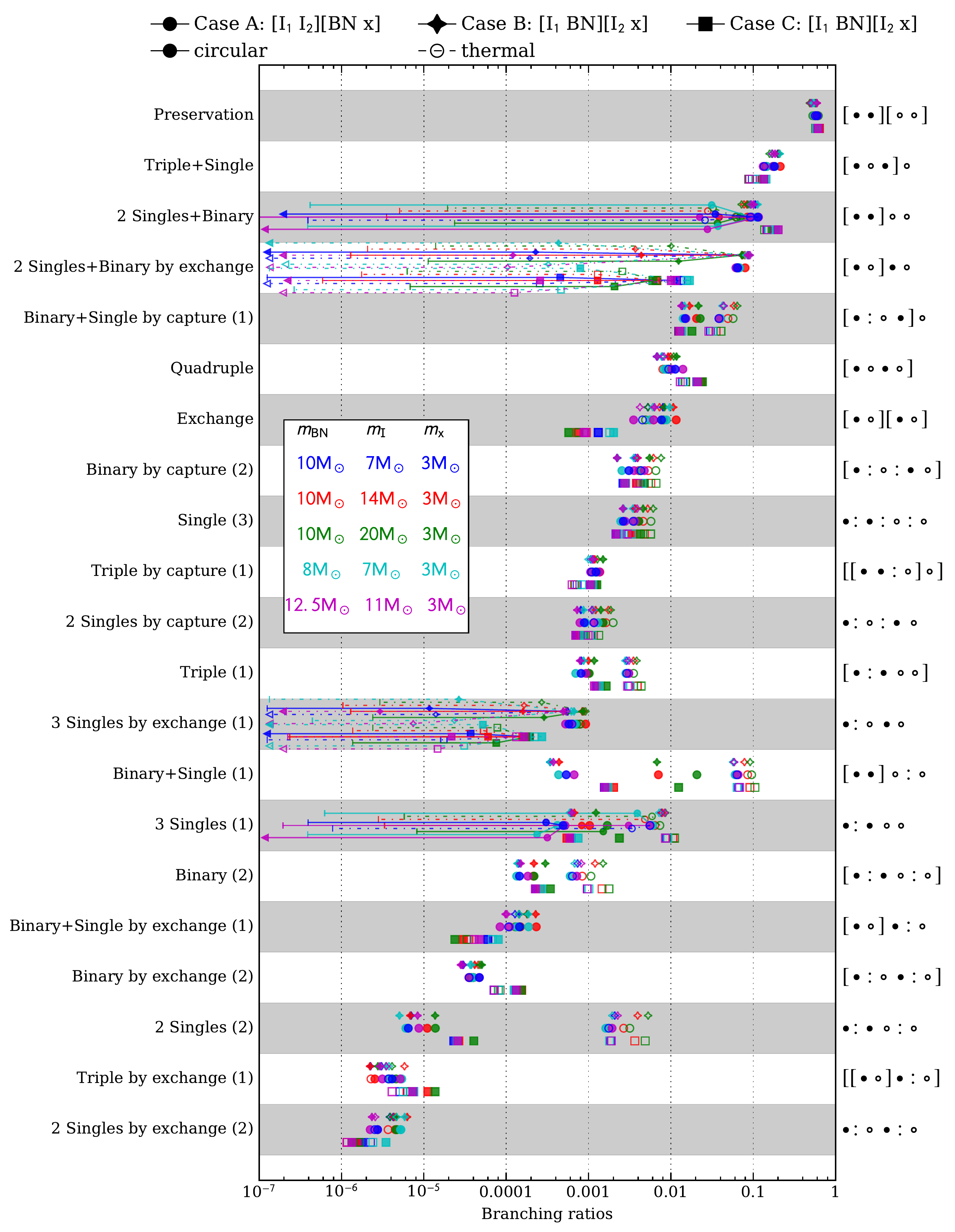}
\caption{
Branching ratios collected for all possible outcomes and experiments carried out in this
work.  Results are sorted from the most to the least probable in the fiducial case (blue
filled circle).  Different symbols represent the different initial configurations: Case A
(circles); Case B (diamonds); and Case C (squares).  Open symbols show cases with thermal
distribution of eccentricities, while filled symbols show where all binaries have
circular orbits initially. Colors show the different assumed masses (see legend). Left
axis labels are the names used to refer to each outcome with the number of collisions
needed for each outcome appearing in parentheses.  Right labels show the schematic
representation of each outcome, similar to \cite{Fregeau2004}, but we distinguish between
exchange of members or preservation of membership of the original binaries.  Branching
ratios of cases that match the BN-x-I observed configuration are a subset of some of the
listed outcomes, these are connected to their parent outcome by a line (dashed for open
symbols, solid for filled symbols).  We mark outcomes where no matching velocity was
found with a left triangle denoting the upper limit of their branching ratio.  Thus only
the outcomes that contain horizontal lines have some chance of producing the observed
BN-x-I system, although typically we only have upper limits on the branching ratio that
leads to the observed system.}
\label{fig:br}
\end{figure*}

By calculating interaction cross sections for the \bnx case we have
obtained a large number of interactions on the order of a few millions
per simulation set with impact parameters from near zero to as large
thousands of AU, by which point no interesting interaction
happens. With this large set of simulations we can calculate branching
ratios of rare interactions down to
$\sim10^{-6}$.  We have classified the outcomes of each experiment
using a similar classification as \cite{Fregeau2004}, however we
distinguish the cases where there was an exchange of members or a
capture of one of the members by one of the binary systems.  Figure
\ref{fig:br} shows all the possible outcomes in this study sorted from
the most to the least probable in our fiducial case with circular
orbits. This trend is similar for the setups with thermal
eccentricities. Figure \ref{fig:br} shows that these models with
thermal eccentricities have a greater chance to obtain a merger, but
not in outcomes that could form the observed BN-x-I system.

In all setups, the most probable case is preservation {i.e., the
  configuration of each setup does not change,} happening $\sim50\%$
of the time, followed by the ``Triple+Single'' case, i.e., the
formation of a stable triple by capturing one member of the other
binary ($\sim20\%$). Our case of interest for Case A, i.e., ``2
Singles+Binary'', comes in third place happening $\sim10\%$ of the
time.  $\outcome$ is a subset of this case. The branching ratio of
this specific subset is marked with the same symbol connected by a
line to the parent set in Figure \ref{fig:br}. The end of the line
shows the branching ratio of the subset that also match the observed
velocities within $2\sigma$.  A left triangle marking the end of the
line means that we did not find any velocity match for this case and
the branching ratio is smaller than the position of the symbol.  The
branching ratio of $\outcome$ is quite high ($~\sim4.5\%$), this means
that the case where the most massive star (BN) is ejected is not so
rare.  However, it is almost impossible to match the observed
velocities in this particular case with the masses assumed by
\cite{Luhman2017}. Increasing the mass of source I considerably
improves the chances to obtain the observed velocities, going from a
branching ratio of $<4\times10^{-7}$ i.e., not a single case with
$m_{\rm{I}}=7\:\MSun$ to a branching ratio of $\sim2\times10^{-5}$ for
$m_{\rm{I}}=20\:\MSun$ (with $m_{\rm{I_1}}=17.14\:\MSun$ and
$m_{\rm{I_2}}=2.86\:\MSun$).

The matching outcome for Cases B and C without a merger involves an
exchange of members, which makes the outcome less frequent, but still
comparable with the original case. The mass of source I also
influences the branching ratios of the matching velocity outcomes,
favoring the cases where source I is more massive.

This trend also remains the same when considering a merger between
I$_1$ and I$_2$, see the ``3 Singles'' case in Fig.~\ref{fig:br}. Even
though the branching ratios of these cases are smaller, the chances of
obtaining the matching velocities are higher since it is the extreme
case where most of the potential energy stored by the binary is
released to the system members. How much energy is set by the
semimajor axis and also individual radius of the source I original
stars. {For this situation,} the radii of the protostars, parameterized
by the factor $\eta$, is one of the largest unknowns in the system and
one of the most important, since it sets the upper limit on the amount
of energy the source I merger can provide.

\end{appendix}

\end{document}